\documentstyle[11pt,paspconf,psfig]{article}

\begin{document}

\title{NICMOS and WFPC2 Imaging of Ultraluminous Galaxies}

\author{
K. D. Borne\affil{Raytheon STX Corporation, NASA Goddard Space Flight Center}}
\author{H.Bushouse, L.Colina, R.A.Lucas, A.Baker, D.Clements,
A.Lawrence, S.Oliver, \& M.Rowan-Robinson}

\begin{abstract}
HST is used to study the power sources and the interaction-induced tidal
disturbances within the most luminous 
galaxies in the local universe --- the Ultra-Luminous IR Galaxies (ULIRGs) ---
through the use of I-band images with WFPC2
and H-band images with NICMOS.
\end{abstract}

% Keywords should be included, but they are not printed in the hardcopy.

\keywords{interacting galaxies, ultraluminous galaxies, starbursts, AGN, HST}

\section{Introduction to Interactive Galaxies and ULIRGs} \label{sec1}

Ultraluminous IR galaxies (ULIRGs; with luminosities
[8--1000$\mu$m] in excess of $10^{12}L_\odot$) are the most luminous
galaxies in the local universe.  They are believed to be
powered by massive bursts of star formation that are induced by
violent galaxy-galaxy collisions (Genzel et al.~1998).  
This ``interaction-activity connection''
and the corresponding physical processes seen in ULIRGs
locally are believed to have been more prevalent at
high redshift -- thus perhaps explaining the quasar phenomenon and maybe
even galaxy formation itself
(Sanders \& Mirabel 1996; and references therein).   
We are obtaining high-resolution images of ULIRGs with the
Hubble Space Telescope (HST).  Such images are probing for the first time
the fine-scale structure in the strong collision-disturbed morphologies 
of these rare and exotic galaxies (Borne et al.~1997a,b,c,d).
The significance of ULIRGs is underscored
by their relevance to galaxy formation, star formation, galaxy evolution,
and the IR background, as evidenced by their prominence within
the overall science missions of ISO, WIRE, SIRTF, and NGST
(e.g., Bushouse et al.~1997).
Their significance
has been further highlighted recently by the
apparent resolution of $\sim$30-50\%
of the IR background in the submillimeter with SCUBA
and the subsequent
identification of those sources with
high-redshift ULIRGs
(Hughes et al.~1998; Barger et al.~1998).

\section{HST Observations and Results} \label{sec2}

We are carrying out a thorough survey and analysis of HST imaging
data on ULIRGs, including nearly 120
$I$-band and over 30 $H$-band images.
All our observations were obtained as part of HST snapshot survey
programs.  We selected objects from the bright
samples of Sanders et al.~(1988a,b) and Melnick \& Mirabel (1990), 
and from the QDOT sample (Leech~et al.~1994;
Lawrence~et al., in press).
The WFPC2 images were obtained 
with the $I$-band filter (F814W at 8000A), with galaxies 
typically centered
in the WF3 chip (800x800 pixels, at 0.1$''$/pixel).
The NICMOS exposures
were obtained with the $H$-band filter (F160W at 1.6$\mu$m).
The images were spiral-dithered in `multiaccum' mode, with
galaxies centered in the NIC2 camera FOV (256x256 pixels,
at 0.075$''$/pixel).

\subsection{WFPC2 Results} \label{subsec2.1}

\begin{figure}
\centerline{\psfig{file=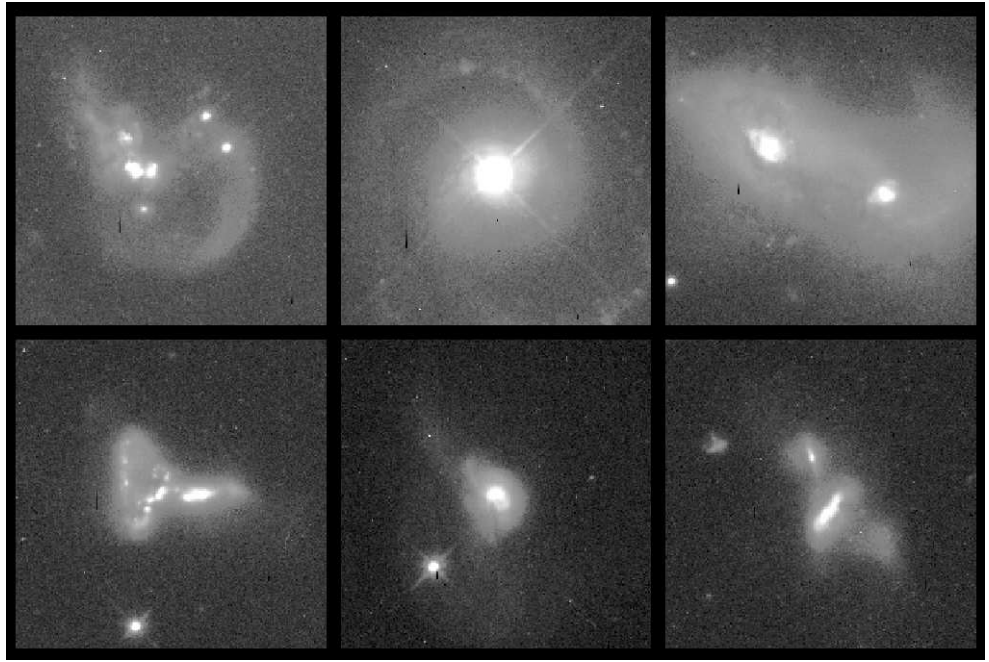,height=8.75cm,width=13.0cm,angle=-90}}
\caption{
Representative WFPC2 images (see text for details).
} \label{figure1}
\end{figure}

We show representative results from our WFPC2 imaging sample in 
Figure~\ref{figure1}.  Each of the 120 ULIRGs that we have
imaged with WFPC2 falls into one of the classes that are
represented by these six examples.
{\it Top Left} -- In nearly all ULIRGs, massive
star formation is seen on all scales in the HST images, including super 
star clusters of the type seen in HST images of other
colliding galaxies, indicating that a giant
starburst is the dominant power source in most ULIRGs.
{\it Top Middle} -- A star--like nucleus is seen in
10\% of the ULIRGs, for which the dominant power source may be a
dust--enshrouded AGN/QSO.
{\it Top Right} --
Several of the galaxies show clear evidence in HST images for
a ring around the central nucleus, very similar to rings seen in HST images
around the centers of other `black hole'--powered galaxies.
{\it Lower Left} --
There is evidence for at least one classical
collisional ring galaxy in the sample, similar to the Cartwheel ring
galaxy imaged by us with HST.
{\it Lower Middle} --
Some ULIRGs previously classified as
non-interacting from ground-based images now show in HST images
clear evidence of merging (a second nucleus) or 
of interaction (e.g., tidal tails).
{\it Lower Right} --
Many ULIRGs appear to have 
physically associated companions, perhaps related to the
collision, merger, and subsequent starburst.

\subsection{NICMOS-WFPC2 Intercomparisons} \label{subsec2.2}

\begin{figure}
\centerline{\psfig{file=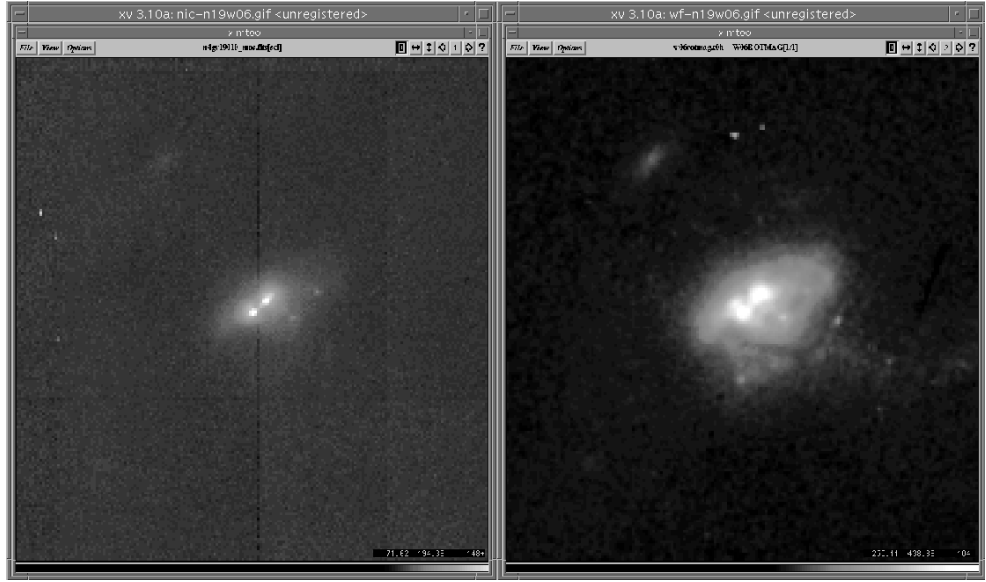,height=7.7cm,width=13.0cm,angle=-90}}
\caption{HST comparison images of IR11095-0238.
Left: NICMOS $H$-band image. 
Right: WFPC2 $I$-band image. (See text for details.)}
\label{figure2}
\end{figure}

We are using our multiple-waveband data set to derive 
wavelength dependencies in the morphologies of ULIRGs, 
hence measuring the spatial variations and effects
of dust, star formation, and stellar population within
these highly disturbed galaxies.  We present some preliminary
comparisons between the NICMOS and WFPC2 images here for several
ULIRGs (listed in Table~\ref{table1}).
A representative example (IR11095-0238) 
is displayed in Figure~\ref{figure2}.

\begin{table}
\caption{Subsample of ULIRGs for NIC--WF Intercomparisons.}
\label{table1}
\begin{center}
\begin{tabular}{llrll}
\tableline
Target Name  &  Redshift  & 
F(60$\mu$m)  &  $\log L_{IR}/L_\odot$  &  AGN ? \\
\tableline
IR11095-0238 ({\it{Fig.~2}}) & 0.1066  & 3.2 Jy  & 12.15  & Liner \\
IR23128-5919                 & 0.0446  & 10.8 Jy & 11.95  & SB $+$ Sey 2(?) \\
QDOT044123+260833            & 0.1712  & 0.8 Jy  & 12.16  & Sey 2 \\
QDOT052321-233443            & 0.1717  & 0.5 Jy  & 12.16  & Sey 2 \\
QDOT062652+350957            & 0.1698  & 0.9 Jy  & 12.22  & \\
QDOT105551+384511            & 0.2066  & 0.6 Jy  & 12.07  & \\
QDOT200342-154747            & 0.1919  & 1.6 Jy  & 12.59  & Sey 1 \\
\tableline
\tableline
\end{tabular}
\end{center}
\end{table}

{\underline{IR11095-0238}} : 
(Displayed in Figure~\ref{figure2}.)
Significant dust, star-forming knots, and tidal
features are seen in the WF image.  Simple double-galaxy morphology
is seen in NIC image, with little dust mottling and weak tidal
features.  Two faint knots
parallel to and just to the right of the two nuclei
are seen in both images.

{\underline{IR23128-5919}} : 
Many knots, small-scale features, and large-scale
features (probably tidal in origin) are seen in
both the NIC and the WF images. 
The appearance of the galaxies 
in the WF image is strongly affected by dust
obscuration, some of which is also seen in the NIC image.

{\underline{QDOT044123+260833}} : 
NIC image shows brighter more compact nucleus.
WF image shows sharper spiral arms, and possibly knots.
The central bar is more prominent in the NIC image.

{\underline{QDOT052321-233443}} : 
WF image shows slightly more extended nucleus.
Tidal arm(s) are visible in both, but very weak
in NIC image.  Tidal arm seen clearly in WF image
of companion.

{\underline{QDOT062652+350957}} : 
Similar appearance in both images, though higher $S/N$
in WF allows for more features to be seen.  
Some possible star-forming knots (tidal dwarf galaxies?) 
are seen clearly in WF image of tidal tails.

{\underline{QDOT105551+384511}} : 
Very different appearance between NIC and WF.
Position angle between two major nuclei differs in
the two images.  Much greater tidal extensions
are seen in the WF image (possible $S/N$ effect).

{\underline{QDOT200342-154747}} : 
Seyfert 1 nucleus dominates, especially in NIC image.
Much more structure seen in WF image: in the
``companion'' and in the surrounding tidal debris. 
Possibly multiple companions are seen in the WF image --- may be
tidal dwarf galaxies in formation.

\acknowledgments

Support for this work was provided by NASA through
grant numbers GO--6346.01--95A and 
GO--7896.01--96A from the Space Telescope Science Institute, which
is operated by AURA, Inc., under NASA contract NAS5--26555.

\end{document}